\documentclass[12pt, onecolumn,draftcls]{IEEEtran}

\usepackage{cite}
\usepackage{graphicx}
\usepackage{amssymb,amsmath}
\usepackage{amsfonts}
\usepackage{multirow}
\usepackage{mathrsfs}
\usepackage{color}

\usepackage{algorithm}
\usepackage{algorithmic}
\usepackage{multirow}
\usepackage{amsmath}

\usepackage{graphics}
\usepackage{graphicx}
\usepackage{epsfig}
\usepackage{cases}
\usepackage{amsmath,bm}
\usepackage{wasysym}
\usepackage{subfigure}

\allowdisplaybreaks

\title{Network-Wide Distributed Carrier Frequency Offsets Estimation and Compensation}
\author{Jian Du and Yik-Chung Wu
\thanks{The authors are with the Department of Electrical and Electronic Engineering, The University of Hong Kong, Pokfulam Road, Hong Kong (e-mail: dujianeee@gmail.com, ycwu@eee.hku.hk).}}

\begin{document}

\newcommand*{\QEDA}{\hfill\ensuremath{\blacksquare}}  
\def\arrow{{\rightarrow}}

\def\N{{\mathcal{N}}}

\def\B{{\mathcal{B}}}
\def\I{{\mathcal{I}}}

\def\diag{{\textrm{diag}}}

\def\i {{ -i}}

\maketitle

\begin{abstract}
In this paper, we propose a fully distributed algorithm for frequency offsets estimation in decentralized systems.
With the proposed algorithm, each node estimates its frequency offsets by local computations and limited exchange of information with its direct neighbors.
Such algorithm does not require any centralized
information processing or knowledge of global network topology.
It is shown analytically that the proposed algorithm always converges to the optimal estimates regardless of network topology.
Simulation results demonstrate the fast convergence of the algorithm and
show that estimation mean-squared-error at each node touches the centralized Cram\'{e}r-Rao bound within a few
{iterations of message exchange}.
Therefore, the proposed method has low overhead and is scalable with network size.
\end{abstract}

\begin{keywords}
Carrier frequency offsets (CFOs) estimation, heterogeneous networks, factor graph, convergence analyses.
\end{keywords}


\newpage
\section{Introduction}\label{Section 1}
In wireless communication systems, local oscillators are used in transceivers to generate carrier signals required for  up-conversion and down-conversion.
Ideally, carrier frequencies produced by oscillators of each transceiver pair should be the same.
However, in practice, frequencies synthesized from independent oscillators will be different from each other due to variation of oscillator circuits.
The received signal impaired
by carrier frequency offsets (CFOs) between transmitter and receiver leads to a continuous rotation of  symbol constellation, resulting in
degradation of system capacity and bit error rate (BER) \cite{Meyr, Moeneclaey, Steendam, MinnCapacity}.
Consequently, carrier frequency synchronization has always been a momentous issue in communication systems.

As modern wireless environments become more heterogeneous and decentralized,
mobile terminals in a network engage more and more cooperative communications and distributed computations \cite{Tse}, new scenarios  that require  multiple wireless units to synchronize with each other arise. For example,

\textbf{Distributed beamforming}:
As shown in Fig. 1(a), to improve the range of communications and save  battery power during the transmission, multiple mobile terminals form a virtual antenna array and cooperatively direct a beam in the
desired direction of transmission \cite{ZQLUO}, \cite{Petropulu}.
Since each source node in the distributed beamformer has an independent local
oscillator, common carrier frequency among all transmitters is crucial to ensure that a beam is aimed in the desired direction.


\textbf{Multi-cell cooperation}: In fully frequency reuse cellular systems as depicted in Fig. 1(b), despite different users interfere with each other, multiple base stations could coordinate their coding and decoding.
It was shown that such joint-processing significantly outperforms a network with individual cell processing \cite{Wyner, Fettweis}.
Yet, multiple base-stations cooperation requires frequency synchronization so that
there is no CFO between each pair of communication link \cite{Gesbert}.

\textbf{Heterogenous Networks (HetNets)}:
HetNets have attracted much attention from both industry and academia in the past few years.
As shown in Fig. 1(c), in a 3-tier HetNet, a mobile may wish to be associated with different tier base stations in the uplink and downlink to obtain optimal performance \cite{Andrews1, Andrews2, Debbah}.
However, multi-tier cooperation is possible only when different tiers of networks are frequency-synchronized to each other.

The above examples of network-wide synchronization problem can be summarized and reduced to a multi-node communication systems as shown in Fig. 2.
Despite relative CFO between each pair of nodes can be optimally estimated by existing methods
\cite{Morelli, Mehrpouyan, Pham,Mon-OnPun, Stoica, GhoghoMIMO, JWChen,Morelli_Interf, Serpedin, FFGao},
network-wide CFOs correction is difficult
since each node needs to synchronize with multiple neighboring nodes with different relative CFOs at the same time.
Making the problem more challenging is the fact that synchronization should be accomplished by local operations without knowing the global network structure since users move around and join different parts of the network randomly.

Pioneering works for network CFOs correction have been proposed in \cite{Mitran},\cite{Zarikoff}.
By gathering all the information in a central processing unit, CFOs are estimated at the receiver and then fedback to corresponding transmitters to adjust the offsets.
These methods are centralized, which are not suitable for large-scale network.
On the other hand,
\cite{MastSlave, RoundTrip, TwoWay} investigated methods for frequency synchronization in distributed beamforming systems.
However, these methods require the formation and maintenance of special network structures (e.g., tree structure in \cite{MastSlave}, ring structure in \cite{RoundTrip} and chain structure in \cite{TwoWay}), thus
suffer from large overhead and long delay, and are not scalable with network size.
Recently, \cite{Spagnolini} proposed the distributed frequency-locked loops (D-FLL) to control and synchronize the carrier frequencies of autonomous nodes based on average consensus principle.
Notwithstanding the distributed carrier frequency calibration advantage, fully distributed D-FLL approach suffers from very slow convergence rate as shown in \cite{Spagnolini}.
Furthermore, the D-FLL algorithm is designed exclusively for single path channel.
Even in a simple point-to-point case with multi-path channel, the D-FLL cannot be applied directly.

In this paper, we propose a network-wide fully distributed CFOs estimation and compensation method which only involves local processing and information exchange between neighboring nodes.
There is no need to have a central hub that aggregates information
and no knowledge about the global network topology is required.
The frequency offset of each oscillator is estimated and corrected locally in each node.
After synchronization, there is no frequency offset between any pair of nodes in the network.
The proposed  algorithm is scalable with network size, and robust to topology changes.
The convergence of the proposed method is also formally proved.

The rest of this paper is organized as follows.
System model is presented in Section \ref{Section 2}.
Fully distributed frequency offsets estimation and correction based on belief propagation (BP) is derived in Section \ref{Section 3} .
The convergence property of the proposed method is analyzed in Section  \ref{Section 4}.
Simulation results are given in Section \ref{Section 4} and, finally, conclusions are provided in Section \ref{Section 5}.

\textit{Notations}: Boldface uppercase and lowercase letters will be used for matrices and vectors, respectively.
$\mathbb{E}$ denotes the expectation over the random variables.
Superscripts $H$ and $T$ denote Hermitian and transpose, respectively.
The symbol $\bm I_N$ represents the $N\times N$ identity matrix,
while $\bm 1_K$ is an all one $K$ dimensional vector.
The symbol $\otimes$ denotes the Kronecker product
and $\odot$ denotes the Hadamard product.
Notation $\N(\bm x|\bm \mu, \bm R)$ stands for the probability density function (pdf) of a Gaussian random vector $\bm x$ with mean $\bm \mu$ and covariance matrix $\bm R$.
The symbol $ \propto$ represents the linear scalar relationship between two real valued functions.
$\mathrm{diag}\{[a_1, \ldots, a_N]\}$ corresponds to an $N\times N$ diagonal matrix with diagonal components $a_1$ through $a_N$, while $\mathrm{blkdiag}\{[\bm A_1, \ldots, \bm A_N]\}$ corresponds to a block diagonal matrix with $\bm A_1$ through $\bm A_N$
as diagonal blocks.
For two matrices  $\bm X$ and $\bm Y$, $\bm X \succeq\bm Y$ means that $\bm X - \bm Y$ is a positive semi-definite matrix.

\section{System Model}\label{Section 2}
We consider a network consisting of $K$ nodes distributed in a field as shown in Fig. \ref{topology}.
The topology of the network is described by a communication graph $\mathcal{G}=(\mathcal{V},\mathcal{E})$ of
order $K$, where $\mathcal{V}=\{1,\ldots, K\}$ is the set of graph
vertexes, and $\mathcal{E}\subseteq \mathcal{V}\times \mathcal{V}$ is the set of graph edges.
In the example shown in Fig. 2, the vertices are depicted by circles and the edges by lines connecting these circles.
The neighborhood of node $i$ is the set of nodes $\mathcal{I}(i)\subset \mathcal{V}$ defined as
$\mathcal{I}(i)\triangleq \{j\in \mathcal{V}| \{i,j\}\in \mathcal{E}\}$, i.e., those nodes that are connected via a direct communication link to node $i$.
It is also assumed that any two distinct nodes can communicate with each other through finite hops, such graph is named strongly connected graph.

In general, relative CFOs exist between any pair of neighboring nodes, and can be estimated by traditional CFOs estimation methods.
Let nodes $i$ and $j$ equipped with $N_i$ and $N_j$
antennas, respectively.
Denote the frequency offsets (with respect to a reference frequency) of the $q^{th}$ antenna on node $i$
as $\omega^{i}_q$, while that of $k^{th}$ antenna of node $j$ as $\omega^{j}_k$.
Then, the relative CFO between the $q^{th}$ and $k^{th}$ antenna of node $i$ and $j$ respectively
is $\epsilon^{i,j}_{q,k}\triangleq\omega^{i}_q - \omega^{j}_k$.
Here we consider the general case where each antenna can be associated with separate oscillator circuit.
Therefore, for the Multiple Input Multiple Output (MIMO) system between node $i$ and node $j$, there are $N_iN_j$ relative CFOs denoted as
$\bm\epsilon^{i,j}\triangleq
[ \epsilon^{i,j}_{1,1},\dots,\epsilon^{i,j}_{N_i,1},
\ldots,
\epsilon^{i,j}_{1,N_j},\dots,\epsilon^{i,j}_{N_i,N_j}]^T$.
Such relative CFOs estimation in MIMO systems can be decomposed into $N_j$ parallel Multiple Input Single Output (MISO)
CFOs estimation problem \cite{Stoica}.
For example, considering a flat-fading MISO system,
for the $k^{th}$ receive antenna of node $j$, the received signal can be written as
\begin{equation} \label{scalorobservation}
y^{i,j}_{k}(t) =\sum_{q=1}^{N_i}h^{ i,j }_{q,k}e^{\jmath \varepsilon^{ i,j }_{q,k} t}z^{i}_{q}(t) + \xi^{j}_{k}(t)
\quad   t=1,\ldots, N,
\end{equation}
where  $h^{ i,j }_{q,k}$ are the unknown channel gains between the $q^{th}$ antenna of node $i$ and $k^{th}$ antenna of node $j$;  $\jmath\triangleq \sqrt{-1}$;
$\{z^{ i}_{q}(t)\}^N_{t=1}$ is the training sequence transmitted from the $q^{th}$ antennas of node $i$;
and $\xi_{k}^{j}(t)$ is the observation noise at the $k^{th}$ antenna of node $j$.
By stacking (\ref{scalorobservation}) with $t=1,\ldots,N$ in vector form and omitting superscript $i,j$ without confusion,
the received vector $\bm y_k \triangleq[y_k(1),\ldots,y_k(N)]^T$ can be written as
\begin{equation}  \label{observation}
\bm y_k = {\bm{ \mathit\Gamma}}_{k}(\bm\epsilon_k) \odot\bm Z_k \bm h_k + \bm\xi_k
\quad   k=1,\ldots, N_j,
\end{equation}
where ${\bm{ \mathit\Gamma}}_k(\bm\epsilon_k) $ is an $N$-by-$N_i$ Vandermonde matrix with its $t^{th}$ row given by
$[e^{\jmath t \epsilon_{1,k}}, e^{\jmath t \epsilon_{2,k}}, \cdots, e^{\jmath t \epsilon_{N_i,k}}]$;
$\bm Z_k$ is the $N$-by-$N_i$ training sequence matrix with its $t^{th}$ row
$[z_1(t), z_2(t), \cdots, z_{N_i}(t)]$;
and $\bm\xi_k=[\xi_k(1),\ldots, \xi_k(N)]^T$ is the observation noise.
The parameters $\bm\epsilon_k\triangleq
[ \epsilon_{1,k}, \epsilon_{2,k},\dots, \epsilon_{N_i,k}]^T$ and $\bm h_k\triangleq
[ h_{1,k},\dots, h_{N_i,k}]^T$ are the parameters need to be estimated.

If the noise is white and Gaussian, i.e.,
$\bm\xi_k\sim \mathcal{CN}(\bm\xi_k;\bm 0,\sigma^2_k \bm I_{N})$,
joint relative CFOs and channels estimation have been extensively studied in the past two decades and the optimal estimates
${\hat{\bm\epsilon}}_{k}$ and $\hat{\bm {h}}_{k}$  have been proposed in\cite{GhoghoMIMO, Stoica, Mon-OnPun, Pham ,JWChen},
with the mean-square-errors (MSEs) approaching the corresponding Cram\'{e}r-Rao bounds (CRBs) in medium and high signal-to-noise ratio (SNR) ranges.
From (\ref{observation}), the CRB of ${\bm\epsilon}_{k}$ can be shown to be \cite{Stoica}
\begin{equation} \label{CRB}
\bm{B}_{{\bm\epsilon}_{k}}({\bm\epsilon}_{k},\bm h_{k})
=\frac{\sigma_k^2}{2} \big\{\mathrm{Re}[\bm V_{k}-\bm T_{k}^H(\bm {\mathit\Lambda}^H_{{k}}\bm {\mathit\Lambda}_{{k}})^{-1}\bm T_{k}]\big\}^{-1},
\end{equation}
where
$\bm V_{k} \triangleq \mathrm{diag}\{\bm h_{k}\}\bm {\mathit\Lambda}^H_{{k}}\bm D^2 \bm {\mathit\Lambda}_{{k}}\mathrm{diag}\{\bm h_{k}\}$,
$\bm T_{k} \triangleq \bm {\mathit\Lambda}^H_{{k}}\bm  D \bm {\mathit\Lambda}_{{k}}\mathrm{diag}\{\bm h_{k}\}$, with
$\bm {\mathit\Lambda}_{{k}}\triangleq {\bm{ \mathit\Gamma}}_k(\bm\epsilon_k)\odot\bm Z_k $
and
$\bm D\triangleq \mathrm{diag}\{[1,2,\ldots, N]\}$.
Since there are $N_j$ independent MISO estimation problems as in (\ref{observation}),
the CRB for frequency estimation in MIMO system between node $i$ and $j$ is given by
$\bm B_{\bm{\epsilon}}^{\{i,j\}}(\{{\bm{\epsilon}}_{k}\}_{k=1}^{N_j},\{{\bm{h}}_{k}\}_{k=1}^{N_j}) = \mathrm{blkdiag}\{[\bm{B}_{{\bm{\epsilon}}_{1}}({\bm{\epsilon}}_{1},\bm {h}_{1}), \ldots, \bm{B}_{{\bm{\epsilon}}_{N_j}}({\bm{\epsilon}}_{N_j},\bm {h}_{N_j})]\}$.


After joint estimation of relative CFOs and channels, the relative CFOs between node $i$ and $j$ can be obtained as
\begin{equation} \label{linearCFO}
\bm r_{i,j} = \bm A_{i,j}\bm \omega_i + \bm A_{j,i}\bm \omega_j + \bm n_{i,j},
\end{equation}
where $\bm r_{i,j} \triangleq [\hat{\bm\epsilon}^T_1,\hat{\bm\epsilon}^T_2,\ldots,\hat{\bm\epsilon}^T_{N_j}]^T $
is the $N_iN_j$ relative CFOs estimates;
$\bm A_{i,j}\triangleq \bm I_{N_i}\otimes \bm 1_{N_j}$
and $\bm A_{j,i}\triangleq - \bm 1_{N_i}\otimes \bm I_{N_j}$;
and $\bm n_{i,j}$ is the estimation error.
It is known that for the maximum likelihood (ML) estimates, ${\bm r}_{i,j}$ is asymptotically
Gaussian distributed with mean
$[{\bm\epsilon}^T_1,{\bm\epsilon}^T_2,\ldots,{\bm\epsilon}^T_{N_j}]^T
=\bm A_{i,j}\bm \omega_i + \bm A_{j,i}\bm \omega_j$
and covariance matrix
equals to
$\bm B_{\bm{\epsilon}}^{\{i,j\}}(\{{\bm{\epsilon}}_{k}\}_{k=1}^{N_j},\{{\bm{h}}_{k}\}_{k=1}^{N_j})$ \cite{Kay}.
That is,
${\bm r}_{i,j} \sim \N({\bm r}_{i,j};{\bm\epsilon}_{i,j},
\bm B^{\{i,j\}}_{\bm{\epsilon}}(\{{{\bm{\epsilon}}}_{k}\}_{k=1}^{N_j},\{{{\bm{h}}}_{k}\}_{k=1}^{N_j})
)$.
Notice that the CRB depends on the true value of $\{{{\bm{\epsilon}}}_{k}\}_{k=1}^{N_j}$ and $\{\bm {h}_{k}\}_{k=1}^{N_j}$
, but since we have obtained the ML estimate $\{{\hat{\bm\epsilon}}\}_{k=1}^{N_j}$ and $\{{{\hat{\bm{h}}}}_{k}\}_{k=1}^{N_j}$,
$\bm B_{\bm{\epsilon}}^{\{i,j\}}(\{{\bm{\epsilon}}_{k}\}_{k=1}^{N_j},\{{\bm{h}}_{k}\}_{k=1}^{N_j})$ can be closely approximated by
$\bm R_{i,j} =
\bm B_{\bm{\epsilon}}^{\{i,j\}}(\{\hat{{\bm{\epsilon}}}_{k}\}_{k=1}^{N_j},\{\hat{{\bm{h}}}_{k}\}_{k=1}^{N_j})$.

With local information (\ref{linearCFO}), the goal is to establish global frequency synchronization.
That is, to  estimate and compensate $\bm \omega_i$ in each node based on estimation results of local relative CFOs
{$\bm r_{i,j}$}.
\newtheorem{remark}{Remark}
\begin{remark}
The system model (\ref{observation}) can be extended to the cases where signals undergoing frequency selective fading channel and even doubly selective channel.
Effective estimators have been extensively studied and MSE performance of these estimators were shown to  touch the corresponding CRBs \cite{Stoica2, Helen10, GhoghoEM}. Thus, we can always establish the relative CFOs relationship as in (\ref{linearCFO}).
\end{remark}
\begin{remark}
 After relative CFOs estimation,  each receiver (node $j$ in the example)
obtains the estimate $\bm r_{i,j}$ as well as the covariance matrix $\bm R_{i,j}$.
By feeding back this information to the corresponding transmitter, node $i$ also obtains the relative CFOs estimates and estimation error covariance.
\end{remark}
\section{Distributed CFOs Estimation}\label{Section 3}

\subsection{Distributed CFOs Estimation via Belief Propagation}
The optimal CFO estimator at each node
in the Bayesian sense
is the minimum mean square error (MMSE) estimator,
which finds the mean of the marginalized posterior distribution:
\begin{equation}\label{MMSE}
\hat{\bm \omega}^{\text{MMSE}}_i\triangleq
\int \cdots \int \bm \omega_i
p\big(\bm \omega_1, \bm \omega_2, \ldots, \bm \omega_K |{\{{\bm r}_{i, j}\}}_{\{i,j\}\in \mathcal{E}}\big) d\bm \omega_2\cdots d\bm \omega_K.
\end{equation}
Here, without loss of generality, node 1 is assumed to be the reference node, so $\bm \omega_1$ is not included in the marginalization.
By using Bayes' theorem, the joint posterior distribution of all frequency offsets can be expressed as:
\begin{eqnarray}\label{joint}
p\big(\bm \omega_1, \bm \omega_2, \ldots, \bm \omega_K|{\{{\bm r}_{i, j}\}}_{\{i,j\}\in \mathcal{E}}\big)
\propto
\prod_{i\in \mathcal{V}} p(\bm\omega_i)
\prod_{{\{i,j\}\in \mathcal{E}}}
 p(\bm r_{i, j}|\bm \omega_i, \bm \omega_j),
\end{eqnarray}
where
$p(\bm\omega_i)$ is the prior distribution
and $p(\bm r_{i, j}|\bm \omega_i, \bm \omega_j)\sim \N(\bm r_{i, j};\bm A_{i, j}\bm\omega_i+\bm A_{j,i}\bm\omega_j,
\bm R_{i, j})$ is the likelihood function.

Notice that since the joint posterior  distribution in (\ref{joint}) depends on interactions
among all unknown variables, the computation of ${\hat{\bm \omega}^{MMSE}_i}$ in (\ref{MMSE}) requires gathering of  all  information in a central processing unit.
However, such centralized processing is not favorable in large-scale networks.

In order to compute the optimal estimate  (\ref{MMSE}) in a distributed way,
one can exploit the conditional independence structure of the joint distribution (\ref{joint}),
which is conveniently revealed by factor graph (FG).
FG is an undirected bipartite graphical representation of a joint distribution that unifies direct and undirected graphical models.
An example of FG in the context of network-wide synchronization is shown in Fig. \ref{FG}.
In the FG, there are two distinct kinds of nodes.
One is variable nodes representing
local synchronization parameters $\bm\omega_i$.
If there is a communication link between node $i$ and node $j$, the corresponding variable nodes
$\bm\omega_i$ and $\bm\omega_j$ are linked by the other kind of node, factor node  $f_{i,j}=p(\bm r_{i,j}|\bm \omega_i, \bm \omega_j)$ representing the local likelihood function
\footnote{Note that $f_{i,j}$=$f_{j,i}$.}.
On the other hand, the factor node $f_{i}=p(\bm \omega_i)$ denotes the prior distribution  of frequency offsets of node
$i$, and is connected only to the variable node $\bm \omega_i$.
Note that the FG is bipartite which means neighbors of a factor node must be variable nodes and vice versa.

From the FG, two kinds of messages are passed around:
One is the message from factor node $f$ (likelihood function $f_{i,j}$ or prior distribution $f_i$) to its neighboring variable node $\bm \omega_i$, defined as
the product of the function $f$ with messages received from all neighboring variable nodes except $\bm \omega_i$,
and then marginalized for $\bm \omega_i$ \cite{Kschischang}
\begin{equation}  \label{BPf2vs1}
m^{(l)}_{f \arrow i}(\bm \omega_i)
=  \int \cdots \int
f\times\! \!\! \! \!\! \!
\prod_{\bm \omega_j\in\B(f)\setminus \bm \omega_i} m^{(l-1)}_{j \arrow f}(\bm \omega_j)
d\{\bm \omega_j\}_{\bm \omega_j\in\B(f)\setminus \bm \omega_i} ,
\end{equation}
where $\B(f)$  denotes the set of  variable nodes that are direct neighbors of the factor nodes $f$ on the FG
and ${\B(f)\setminus \bm \omega_i}$ denotes the same set but with $\bm \omega_i$ removed.
In (\ref{BPf2vs1}), $m^{(l-1)}_{j \arrow f}(\bm \omega_j)$ is the other kind of message from variable node to factor node which is simply the product of the incoming messages on other links, i.e.,
\begin{equation} \label{BPvs2f1}
m^{(l)}_{j \arrow f}(\bm \omega_j)
=\prod_{\tilde{f}\in\B(\bm \omega_j)\setminus f} m^{(l)}_{ \tilde{f} \arrow j}
 (\bm \omega_i),
\end{equation}
where $\B(\bm \omega_j)$  denotes the set of  factor nodes that are direct neighbors of the variable nodes $\bm \omega_j$ on the FG.
It can be seen from (\ref{BPf2vs1}) and (\ref{BPvs2f1}) that for both variable nodes and factor nodes, each outgoing message is a function of all
incoming messages in the last round except the incoming message from the node where the outgoing message will
be directed to.
This essential restriction guarantees that for cycle-free FG, incoming and outgoing messages
on each edge are independent, and at the end the algorithm produces correct marginal posterior
distribution.

The two kinds of messages are iteratively updated at variable nodes and factor nodes, respectively.
In any round of message exchange, a belief of $\bm \omega_i$ can be computed at variable node $i$ as the product of all the incoming messages from neighboring factor nodes, which is given by
\begin{equation} \label{BPbelief}
b^{(l)}(\bm \omega_i)
 =  \prod_{ f\in\B(\bm \omega_i)} m^{(l)}_{ f \arrow i}(\bm \omega_i).
\end{equation}
Thereupon, the estimate of $\bm \omega_i$ in the $l^{th}$ iteration is simply
\begin{equation}\label{est1}
\hat {\bm \omega}_i^{(l)}
 =  \int \bm \omega_i b^{(l)}(\bm \omega_i)d\bm \omega_i.
\end{equation}
Notice that after convergence, the belief  $b^{(l)}(\bm \omega_i)$ at each variable node corresponds to the marginal distribution of that variable exactly
when the underlying FG is loop free \cite{Kschischang}.
However, for  the FG with loops, it is generally difficult to known if BP will converge \cite{WalkSum1}.
Even if BP converge to a fixed point, there is no guarantee on the estimation accuracy.
Despite the lack of general results on BP, in this paper, the convergence and optimality of BP for network-wide CFO estimation algorithm will be proved in section \ref{Section 4}.

\subsection{Message Computation}
In the BP framework, messages are passed and updated iteratively.
In order to start the recursion,
in the first round of message passing, it is reasonable to set the initial messages from factor nodes to variable nodes
$m^{(0)}_{f_{i} \arrow  i}(\bm \omega_i)$ and $m^{(0)}_{f_{i,j} \arrow  i}(\bm \omega_i)$ as $p(\bm\omega_i)$ and non-informative message $\mathcal{N}(\bm\omega_i;\bm v^{(0)}_{f_{i,j}\arrow i}, \bm C_{f_{i,j}\arrow i}^{(0)})$, respectively, where
$\bm v^{(0)}_{f_{i,j}\arrow i}$ can be arbitrarily chosen and $[\bm C_{f_{i,j}\arrow i}^{(0)})]^{-1}=\bm 0$.
Assuming $p(\bm \omega_i)=m^{(0)}_{f_{i,j} \arrow  i}(\bm \omega_i)$ is in Gaussian form $ \N(\bm\omega_i; \bm v_i, \bm C_i)$ (if there is no prior information, we can set the mean to be zero and set the variance to be a large value, i.e., non-informative prior).
Thereupon, based on the fact that the likelihood function $f_{i,j}$ is also Gaussian,
according to (\ref{BPf2vs1}),  $m^{(1)}_{f_{i,j} \arrow  i}(\bm \omega_i)$ is a Gaussian
function.
In addition, $m^{(1)}_{j \arrow f_{i,j}}(\bm \omega_j)$ being the product of Gaussian functions in (\ref{BPvs2f1}) is also a Gaussian function \cite{Papoulis}.
Thus during each round of message exchange, all the messages are Gaussian functions and
only the mean vectors and covariance matrices need to be exchanged between factor nodes and variable nodes.

At this point, we can compute the messages at any iteration.
In general, for the $l^{th}$ ($l=2,3,\cdots$) round of message exchange, factor node $f_{i,j}$ receive messages $m^{(l-1)}_{ j \arrow f_{i,j}} (\bm \omega_j)$ from its neighboring variable nodes and then compute messages using (\ref{BPf2vs1}).
After some derivations, it can be obtained that
\begin{eqnarray}\label{BPf2vs2}
m^{(l)}_{f_{i,j} \arrow  i}(\bm \omega_i)
&=&  \int
p(\bm A_{i,j},\bm A_{j,i}|\bm \omega_i, \bm \omega_j)
m^{(l-1)}_{j \arrow f_{i,j} }(\bm \omega_j)
d\bm \omega_j  \nonumber \\
&\propto&
\mathcal{N}(\bm\omega_i;\bm v^{(l)}_{f_{i,j}\arrow i}, \bm C_{f_{i,j}\arrow i}^{(l)}),
\end{eqnarray}
where the inverse of covariance matrix is
\begin{equation}\label{messagecov}
\big[\bm C_{f_{i,j}\arrow i}^{(l)}\big]^{-1}
 =
\bm A_{i,j}^T
\bigg[ \bm R_{i,j}+ \bm A_{j,i}\bm C_{j\arrow f_{i,j}}^{(l-1)} \bm A_{j,i}^T \bigg]^{-1}
\bm A_{i,j},
\end{equation}
and the mean vector is
\begin{equation}\label{f2vm}
\bm v^{(l)}_{f_{i,j}\arrow i}
=
\bm C_{f_{i,j}\arrow i}^{(l)}\bm A_{i,j}^T \bigg[\bm R_{i,j}
+ \bm A_{j,i}\bm C_{j\arrow f_{i,j}}^{(l-1)}\bm A_{j,i}^T \bigg]^{-1}
(\bm r_{i,j}-\bm A_{j,i}\bm v^{(l-1)}_{j\arrow f_{i,j}}).
\end{equation}

On the other hand, using (\ref{BPvs2f1}), the messages passed from variable nodes to factor nodes can be computed as
\begin{eqnarray}
m^{(l)}_{i \arrow f_{i,j}}(\bm \omega_i) \label{BPvs2f2}
& = &
\prod_{f\in\B(\bm \omega_i)\setminus f_{i,j}} m^{(l)}_{ f \arrow i} (\bm \omega_i)
\nonumber \\
& \propto&
\mathcal{N}(\bm\omega_i;\bm v^{(l)}_{i\arrow  f_{i,j}}, \bm C^{(l)}_{i\arrow  f_{i,j}}),
\end{eqnarray}
where
\begin{equation}\label{var2fCov}
\big[\bm C^{(l)}_{i\arrow  f_{i,j}}]^{-1} =  \sum_{f\in\B(\bm \omega_i)\setminus f_{i,j}}
\big[\bm C_{f\arrow i}^{(l)}\big]^{-1},
\end{equation}
and
\begin{equation} \label{v2fm}
\bm v^{(l)}_{i\arrow  f_{i,j}}= \bm C^{(l)}_{i\arrow  f_{i,j}}
 \sum_{f\in\B(\bm \omega_i)\setminus f_{i,j}}
\big[\bm C_{f\arrow i}^{(l)}\big]^{-1}\bm v^{(l)}_{f\arrow i}.
\end{equation}

Furthermore, during each round of message passing, each node can compute the belief for $\bm\omega_i$
using (\ref{BPbelief}), which can be easily shown to be
$b_{i}^{(l)}(\bm\omega_i)\sim  \mathcal{N}(\bm\omega_i;\bm \mu_i^{(l)}, \bm P_{i}^{(l)})$, with the
inverse of covariance matrix
\begin{equation} \label{beliefP}
 \big[\bm P_{i}^{(l)}\big]^{-1} = \sum_{j\in\I(i)}\big[\bm C_{f_{i,j}\arrow i}^{(l)}\big]^{-1},
\end{equation}
and mean vector
\begin{equation}\label{beliefu}
 \bm \mu_i^{(l)} = \bm P_{i}^{(l)} \sum_{j\in\I(i)}\big[\bm C_{f_{i,j}\arrow i}^{(l)}\big]^{-1}\bm v^{(l)}_{f_{i,j}\arrow i}.
\end{equation}
When the algorithm converges or the maximum number of message exchange is reached, each node computes
the CFOs according to (\ref{est1}) as
\begin{equation}\label{est2}
\hat {\bm \omega}_i^{(l)}
 =  \int \bm \omega_i b^{(l)}(\bm \omega_i)d\bm \omega_i
 = \bm \mu_i^{(l)}.
\end{equation}

The iterative algorithm based on BP is summarized as follows.
The algorithm is started by setting the message from factor node to variable node as
$m^{(0)}_{f_{i} \arrow  i}(\bm \omega_i)=p(\bm\omega_i)$ and $m^{(0)}_{f_{i,j} \arrow  i}(\bm \omega_i)=\mathcal{N}(\bm\omega_i;\bm v^{(0)}_{f_{i,j}\arrow i}, \bm C_{f_{i,j}\arrow i}^{(0)})$
with $\bm v^{(0)}_{f_{i,j}\arrow i}=\bm 0$ and $[\bm C_{f_{i,j}\arrow i}^{(0)})]^{-1}=\bm 0$.
At each round of message exchange, every variable node computes the output messages to factor nodes
according to (\ref{var2fCov}) and (\ref{v2fm}).
After receiving the messages from its neighboring variable nodes, each factor node computes its output messages according to (\ref{messagecov}) and (\ref{f2vm}).
Such iteration is terminated when (\ref{beliefu}) converges (e.g., when $\|\mu_i^{(l)}-\mu_i^{(l-1)}\|<\eta$, where $\eta$ is a threshold) or the maximum number of iteration is reached.
Then the estimate of CFOs of each node is obtained as in  (\ref{est2}).

\begin{remark}
 In practical networks, there is neither factor nodes nor variable nodes. The two kinds of messages
$m^{(l)}_{i \arrow f_{i,j}}(\bm \omega_i) $ and $m^{(l)}_{f_{i,j} \arrow j}(\bm \omega_j)$
are computed locally at node $i$, and only mean vector $ \bm v^{(l)}_{f_{i,j} \arrow  j}(\bm \omega_j)$
and covariance matrix
$\bm C_{f_{i,j}\arrow j}^{(l)}(\bm \omega_j)$
are passed from node $i$ to node $j$ during each round of message exchange of BP.
It can be seen the algorithm is fully distributed and each node only needs to exchange limited information with neighboring nodes.
\end{remark}
\begin{remark}
Since each pair of node has knowledge of relative CFOs and channel between them, the BP message exchange can be performed as in point-to-point communications.
\end{remark}
\section{Theoretical Analyses of BP method}\label{Section 4}
It is generally known that if the FG contains cycles, such as the one shown in Fig. \ref{FG},
messages can flow many times around the graph, leading to the possibility of divergence of BP algorithm \cite{DiagnalDominant}.
A general sufficient condition for convergence of loopy FGs is given in
\cite{WalkSum1}.
Unfortunately, it requires the knowledge of the joint posterior
distribution of all unknown variables as shown in (\ref{joint}), and is difficult to verify for large-scale dynamic networks.
Recently, \cite{LengMei11BP} proved the convergence of BP in the context of distributed  clock offset synchronization in wireless sensor network.
The convergence is established for scalar variables in which sub-stochastic and irreducible properties of BP message recursion were exploited.
However, in vector variable case,
the BP messages involve matrix inverses (see  (\ref{messagecov}), (\ref{f2vm}), (\ref{var2fCov}) and (\ref{v2fm})), and the sub-stochastic and irreducible properties cannot be easily applied.
In the following, we will prove the convergence of BP messages in vector form, and show that the BP based CFO estimates asymptotically converge to the optimal MMSE solution regardless of network topology.


{\textbf{Theorem 1.}} The covariance matrix $\bm P_{i}^{(l)}$ of belief $b_i^{(l)}(\bm\omega_i)$ at each node converges, and there exists a positive definite matrix ${\bm P_{i}^{\ast}}$ such that $\lim_{l \to +\infty}\bm P_{i}^{(l)}=\bm P_{i}^{\ast}$ regardless of network topology.

\noindent  \textit{Proof}:
We begin with a few properties of positive semi-definite (p.s.d.) matrices
and positive definite (p.d.) matrices.
If $\bm X $, $\bm Y $, $\bm Z$ are $N_i$-by-$N_i$
matrices and
 $\bm X \succ \bm 0$, $\bm Y \succ \bm 0$, $\bm Z \succeq \bm 0$, then we have

\noindent Property  \romannumeral 1): $\bm X^{-1} \succ 0$.

\noindent Property  \romannumeral 2): $\bm X + \bm Y \succ 0$.

\noindent Property  \romannumeral 3): $\bm X + \bm Z \succ 0$.

\noindent Property  \romannumeral 4): $ \bm X  \succeq  \bm Y $ if and only if
$\bm Y^{-1} \succeq  \bm X^{-1}$ \cite{MatrixTricks}.

\noindent Property  \romannumeral 5):
$\bm A_{i,j}^T\bm X\bm A_{i,j}\succ \bm 0$
and $\bm A_{i,j}^T\bm Z\bm A_{i,j}\succeq \bm 0$, where $\bm A_{i,j}$ is defined in (\ref{linearCFO}).

\noindent Property  \romannumeral 6): $\bm A_{j,i}\bm X\bm A^T_{j,i}\succeq 0$.

\noindent Properties  \romannumeral 1) to \romannumeral 4) are standard results in matrix analysis.  Property \romannumeral 5) is true due to the fact that $\bm A_{i,j}$  is of full column rank.
The proof of \romannumeral 6) follows from the definition of $\bm X \succ \bm 0$ which is $\bm y^T\bm X\bm y \geq 0$ for any $\bm y$ (including all zeros vector).
The result is obtained if we let $\bm y=\bm A_{j,i}^T\bm x$.

Next, we investigate the updating properties of the message covariance matrix.
Substituting (\ref{var2fCov}) into (\ref{messagecov}),
the covariance update rules from factor nodes to variable nodes are
\begin{equation} \label{f2fC}
\big[\bm C_{f_{i,j}\arrow i}^{(l)}\big]^{-1}
 =
\bm A_{i,j}^T
\bigg[ \bm R_{i,j}+ \bm A_{j,i}
\bigg[\bm C_{f_{j}\arrow j}^{-1}
+\sum_{f_{k,j}\in\B(\bm \omega_j)\setminus f_{i,j}} \big[\bm C_{f_{k,j}\arrow j}^{(l-1)}\big]^{-1}\bigg]^{-1}
 \bm A_{j,i}^T \bigg]^{-1}
\bm A_{i,j}.
\end{equation}
From (\ref{f2fC}), we can deduce two consequences.
First, if all message covariance $\bm C_{f_{k,j}\arrow j}^{(l-1)}$
and  prior covariance $\bm C_{f_{j}\arrow j}$ on the right-hand-side of (\ref{f2fC}) are non-informative,
$\big[\bm C_{f_{i,j}\arrow i}^{(l)}\big]^{-1}$ cannot be updated, i.e., $\big[\bm C_{f_{i,j}\arrow i}^{(l)}\big]^{-1}=\bm 0$.
On the other hand, if some of the $[\bm C_{f_{k,j}\arrow j}^{(l-1)}]^{-1}$
or $\bm C_{f_{j}\arrow j}^{-1}$  on the right-hand-side of (\ref{f2fC}) are p.d. while the remaining are $\bm 0$,
then
$\big[\bm C_{f_{j}\arrow j}^{-1}
+\sum_{f_{k,j}\in\B(\bm \omega_j)\setminus f_{i,j}} \big[\bm C_{f_{k,j}\arrow j}^{(l-1)}\big]^{-1}\big]^{-1} \succ \bm 0$ according to property \romannumeral 3).
Applying property \romannumeral 6), we have
$\bm A_{j,i}
\bigg[\bm C_{f_{j}\arrow j}^{-1}
+\sum_{f_{k,j}\in\B(\bm \omega_j)\setminus f_{i,j}} \big[\bm C_{f_{k,j}\arrow j}^{(l-1)}\big]^{-1}\bigg]^{-1}
 \bm A_{j,i}^T \succeq  0$.
Furthermore, since $\bm R_{i,j}$ is the relative CFO estimation covariance, we have  $\bm R_{i,j}\succ \bm 0$.
Thus, $\bm R_{i,j}+ \bm A_{j,i}
\bigg[\bm C_{f_{j}\arrow j}^{-1}
+\sum_{f_{k,j}\in\B(\bm \omega_j)\setminus f_{i,j}} \big[\bm C_{f_{k,j}\arrow j}^{(l-1)}\big]^{-1}\bigg]^{-1}
 \bm A_{j,i}^T \succ \bm 0$.
Then, based on properties \romannumeral 1) and \romannumeral 5),
we obtain $\big[\bm C_{f_{i,j}\arrow i}^{(l)}\big]^{-1} \succ  \bm 0$.
We summarize the above discussion as
\begin{equation} \label{C-evol}
\big[\bm C_{f_{i,j}\arrow i}^{(l)}\big]^{-1}
\left\{
\begin{array}{l}
=\bm 0,  \quad  \textrm{if all $[\bm C_{f_{k,j}\arrow j}^{(l-1)}]^{-1}$ and $\bm C_{f_{j}\arrow j}^{-1}=\bm 0$ }, \\
\succ \bm 0, \quad  \textrm{if some of $[\bm C_{f_{k,j}\arrow j}^{(l-1)}]^{-1}$ or $\bm C_{f_{j}\arrow j}^{-1}\succ \bm 0$, while others are $\bm 0$}.
\end{array}
\right.
\end{equation}

Now we prove that for any node $i$,
if there exists a directed path from node $1\arrow\ldots\arrow j \arrow i$ in the network topology, there must be a finite iteration number $s$ such that
$\big[\bm C_{f_{i,j}\arrow i}^{(s+1)}\big]^{-1}
\succeq
\big[\bm C_{f_{i,j}\arrow i}^{(s)}\big]^{-1}
\succ
\big[\bm C_{f_{i,j}\arrow i}^{(s-1)}\big]^{-1} =\bm 0.$
Initially, all $[\bm C_{f_{k,j}\arrow j}^{(0)}]^{-1}$ and $\bm C_{f_{j}\arrow j}^{-1}$ over the FG equal $\bm 0$
except $\bm C_{f_{1}\arrow 1}^{-1}=\infty \bm I$ at the reference node.
Hence, the message update starts from the reference node.
More explicitly, $\forall i$:
$\{i,1\}\in\mathcal{E}$, the message covariance is obtained by putting $j=1$ into (\ref{f2fC}), which is
\begin{equation} \label{refV}
\big[\bm C_{f_{i,1}\arrow i}^{(1)}\big]^{-1}
= \bm A_{i,1}^T \bm R_{i,1}^{-1}\bm A_{i,1}\succ \bm 0,
\end{equation}
where the p.d. property is due to property \romannumeral 5).
Furthermore, since $\bm C_{f_{1}\arrow 1}^{-1} =\infty \bm I$, it will dominate the sum
$\bm C_{f_{j}\arrow j}^{-1}
+\sum_{f_{k,j}\in\B(\bm \omega_j)\setminus f_{i,j}} \big[\bm C_{f_{k,j}\arrow j}^{(l-1)}\big]^{-1}$
in (\ref{f2fC}),
and lead to $\big[\bm C_{f_{i,1}\arrow i}^{(l)}\big]^{-1}=\bm A_{i,1}^T \bm R_{i,1}^{-1}\bm A_{i,1}$.
Thus, we have
\begin{equation} \label{refV2}
\ldots=\big[\bm C_{f_{i,1}\arrow i}^{(l)}\big]^{-1}=\ldots =\big[\bm C_{f_{i,1}\arrow i}^{(2)}\big]^{-1}=\big[\bm C_{f_{i,1}\arrow i}^{(1)} \big]^{-1}
\succ \big[\bm C_{f_{i,1}\arrow i}^{(0)} \big]^{-1}
=\bm 0.
\end{equation}

Then, we consider all nodes $i$ with a directed path node $1\arrow j \arrow i$.
In the $1^{st}$ iteration, node $j$ will take the position of node $i$ in (\ref{refV2})
implying $\big[\bm C_{f_{j,1}\arrow j}^{(1)} \big]^{-1}\succ \bm 0$, while $\big[\bm C_{f_{i,j}\arrow i}^{(1)} \big]^{-1}$ has not been updated in the first iteration, i.e., $\big[\bm C_{f_{i,j}\arrow i}^{(1)} \big]^{-1}=\bm 0$.
In the second iteration, from (\ref{C-evol}), we have $\big[\bm C_{f_{i,j}\arrow i}^{(2)} \big]^{-1}\succ\bm 0$.
Since $\big[\bm C_{f_{k,j}\arrow j}^{(1)} \big]^{-1} $  and $\bm C_{f_{j}\arrow j}^{-1}  $ equal $\bm 0$, taking inverse on $\bm C_{f_{j}\arrow j}^{-1}
+\sum_{f_{k,j}\in\B(\bm \omega_j)\setminus f_{i,j}} \big[\bm C_{f_{k,j}\arrow j}^{(1)}\big]^{-1}$ gives
$ \bigg[\bm C_{f_{j}\arrow j}^{-1}
+\sum_{f_{k,j}\in\B(\bm \omega_j)\setminus f_{i,j}} \big[\bm C_{f_{k,j}\arrow j}^{(1)}\big]^{-1}\bigg]^{-1}
\succeq
\bigg[\bm C_{f_{j}\arrow j}^{-1}
+\sum_{f_{k,j}\in\B(\bm \omega_j)\setminus f_{i,j}} \big[\bm C_{f_{k,j}\arrow j}^{(2)}\big]^{-1}\bigg]^{-1}$.
Further applying properties \romannumeral 6), \romannumeral 4) and \romannumeral 5), we obtain
\begin{eqnarray} \label{ineq}
&&\underbrace{\bm A_{i,j}^T
\bigg[ \bm R_{i,j}+ \bm A_{j,i}
\bigg[\bm C_{f_{j}\arrow j}^{-1}
+\sum_{f_{k,j}\in\B(\bm \omega_j)\setminus f_{i,j}} \big[\bm C_{f_{k,j}\arrow j}^{(2)}\big]^{-1}\bigg]^{-1}
 \bm A_{j,i}^T \bigg]^{-1}
\bm A_{i,j}}_{=\big[\bm C_{f_{i,j}\arrow i}^{(3)}\big]^{-1}}     \nonumber \\
&\succeq&
\underbrace{\bm A_{i,j}^T
\bigg[ \bm R_{i,j}+ \bm A_{j,i}
\bigg[\bm C_{f_{j}\arrow j}^{-1}
+\sum_{f_{k,j}\in\B(\bm \omega_j)\setminus f_{i,j}} \big[\bm C_{f_{k,j}\arrow j}^{(1)}\big]^{-1}\bigg]^{-1}
 \bm A_{j,i}^T \bigg]^{-1}
\bm A_{i,j}}_{=\big[\bm C_{f_{i,j}\arrow i}^{(2)}\big]^{-1}}.
\end{eqnarray}
Thus $\big[\bm C_{f_{i,j}\arrow i}^{(3)} \big]^{-1}\succeq \big[\bm C_{f_{i,j}\arrow i}^{(2)}\big]^{-1}
\succ \big[\bm C_{f_{i,j}\arrow i}^{(1)} \big]^{-1}=\bm 0$.
In general, for any node $i$, if there exists a directed path from node $1\arrow\ldots\arrow j \arrow i$ in the network topology, there must be a finite iteration number $s$ such that
\begin{equation} \label{Cchain}
\big[\bm C_{f_{i,j}\arrow i}^{(s+1)}\big]^{-1}
\succeq
\big[\bm C_{f_{i,j}\arrow i}^{(s)}\big]^{-1}
\succ
\big[\bm C_{f_{i,j}\arrow i}^{(s-1)}\big]^{-1} =\bm 0.
\end{equation}

Finally, we divide the discussion into three cases, covering all possible relationships between two neighboring node $i$ and $j$:

\noindent a) there exists a path from node $1\arrow\ldots\arrow j \arrow i$  and $ j\neq 1$;

\noindent b) there exists a path from node $1 \arrow i$;

\noindent c) there is no path from node $1\arrow\ldots\arrow j \arrow i$.

\noindent For the first case, suppose
$\big[\bm C_{f_{i,j}\arrow i}^{(l)}\big]^{-1}\succeq \big[\bm C_{f_{i,j}\arrow i}^{(l-1)}\big]^{-1}$ holds
for $l>s$.
Since $j\neq 1$, there must be a node $k$, such that
$\big[\bm C_{f_{k,j}\arrow j}^{(l)}\big]^{-1}\succeq \big[\bm C_{f_{k,j}\arrow j}^{(l-1)}\big]^{-1}$.
Then, it can be easily shown that $ \sum_{f_{k,j}\in\B(\bm \omega_j)\setminus f_{i,j}}\big[\bm C_{f_{k,j}\arrow j}^{(l)}\big]^{-1}
\succeq
\sum_{f_{k,j}\in\B(\bm \omega_j)\setminus f_{i,j}}\big[\bm C_{f_{k,j}\arrow j}^{(l-1)}\big]^{-1} $.
Following the same arguments above (\ref{ineq}), it can be obtained that
$\big[\bm C_{f_{i,j}\arrow i}^{(l+1)}\big]^{-1}\succeq \big[\bm C_{f_{i,j}\arrow i}^{(l)}\big]^{-1}$.
Hence, by induction we have
\begin{equation}\label{Cseq2}
 \big[\bm C_{f_{i,j}\arrow i}^{(l)}\big]^{-1}\succeq \ldots \succeq \big[\bm C_{f_{i,j}\arrow i}^{(s+1)}\big]^{-1} \succeq \big[\bm C_{f_{i,j}\arrow i}^{(s)}\big]^{-1}
\succ \bm 0,
\quad
\textrm{ for $l>s$}.
\end{equation}

\noindent For the second case, if there exists a path node $1 \arrow i$, the corresponding result is in (\ref{refV2}).
For the third case, if the path node $1\arrow\ldots\arrow j \arrow i$ does not exist,
$\big[\bm C_{f_{i,j}\arrow i}^{(l)}\big]^{-1}$ never get update, and always equals to $\big[\bm C_{f_{i,j}\arrow i}^{(0)}\big]^{-1}=\bm 0$.

Since strongly connected network is considered,
there is at least one $\tilde{j}\in\I(i)$ such that the first
case is true, therefore, we obtain
\begin{equation} \label{C1}
\ldots \succeq
\sum_{j\in\I(i)}\big[\bm C_{f_{i,j}\arrow i}^{(l+1)}\big]^{-1}
\succeq
\sum_{j\in\I(i)}\big[\bm C_{f_{i,j}\arrow i}^{(l)}\big]^{-1}
\succeq \ldots
\sum_{j\in\I(i)}\big[\bm C_{f_{i,j}\arrow i}^{(s)}\big]^{-1}
\succ \bm 0,
\quad \textrm{for $l> s$}.
\end{equation}
Applying matrix inverse to (\ref{C1}) and using the definition of $ \bm P_i^{(l)}$ in (\ref{beliefP}), we have
\begin{equation} \label{P1}
 \bm P_i^{(s)}  \succeq \ldots \succeq  \bm P_i^{(l)}   \succeq  \bm P_i^{(l+1)} \succeq \ldots \succ \bm 0 ,
\quad \textrm{for $l> s$},
\end{equation}
where the p.d. property of $ \bm P_i^{(l)}$ is due to property \romannumeral 1).
Consequently such non-increasing p.d. matrix sequence converges to certain $\bm P_{i}^{\ast}\succ \bm 0$ \cite{MatrixConverge}.
\QEDA

The importance of Theorem $1$ is that if a reference node exists, the belief covariance matrices always converge.
Next, we investigate the convergence of belief mean vectors.

{\textbf{Theorem 2.}}
The mean $\bm\mu_i^{(l)}$ of the belief $b^{(l)}(\bm \omega_i)$ converges to a fixed  a vector $\bm \mu_i^{\ast}$ regardless of the network topology, i.e., $\lim_{l \to +\infty} \bm \mu_i^{(l)}= \bm \mu_i^{\ast}$.

\noindent  \textit{Proof}:
From the proof of Theorem 1, there are three cases of relationships between node $i$ and node $j$ (above (\ref{Cseq2})).
For the first and second cases, the evolution of  $\big[\bm C_{f_{i,j}\arrow i}^{(l)}\big]^{-1}$ are described by (\ref{refV2}) and (\ref{Cseq2}), respectively.  Taking matrix inverse of (\ref{refV2}) and (\ref{Cseq2}), we can readily see that $\bm C_{f_{i,j}\arrow i}^{(l)}$ is a monotonically decreasing matrix sequence and bounded below by $\bm 0$.  Thus, $\bm C_{f_{i,j}\arrow i}^{(l)}$ is convergent.
For the third case, $\bm C_{f_{i,j}\arrow i}^{(l)}$ is never updated, and thus can also be viewed as  convergent.  On the other hand, computation of $\bm C_{j\arrow f_{i,j}}^{(l)}$ depends on $\bm C_{f_{i,j}\arrow i}^{(l)}$ as shown in (\ref{var2fCov}).
So, if $\bm C_{f_{i,j}\arrow i}^{(l)}$ is convergent, then $\bm C_{j\arrow f_{i,j}}^{(l)}$  is also convergent.  In this proof, it is assumed that $\bm C_{f_{i,j}\arrow i}^{(l)}$ and $\bm C_{j\arrow f_{i,j}}^{(l)}$ have already converged to $\bm C_{f_{i,j}\arrow i}^{(\ast)}$ and $\bm C_{j\arrow f_{i,j}}^{(\ast)}$, respectively, as the convergence of message covariance matrices do not depend on the message mean vectors.

Substituting (\ref{v2fm}) into (\ref{f2vm}), we obtain the mean update rules from factor nodes to variable nodes
as
\begin{eqnarray}\label{f2f2}
\bm v^{(l)}_{f_{i,j}\arrow i}
&=&\underbrace{
\bm C_{f_{i,j}\arrow i}^{(\ast)}\bm A_{i,j}^T \bigg[\bm R_{i,j}
+ \bm A_{j,i} \bm C_{j\arrow f_{i,j}}^{(\ast)}\bm A_{j,i}^T\bigg]^{-1} }_{\triangleq\bm M_{i,j}}\\\nonumber
&&
\bigg\{\underbrace{\bm r_{i,j} - \bigg[\bm A_{j,i}\bm C^{(\ast)}_{j\arrow f_{i,j}}\bm C_{f_{j}\arrow{j}}^{-1}\bm v_{f_{j}\arrow{j}}}_{\triangleq\bm a_{i,j}}
+\underbrace{\bm A_{j,i}\bm C^{(\ast)}_{j\arrow f_{i,j}}}_{\triangleq\bm F_{i,j}}
 \sum_{f_{j,k}\in\B(\bm \omega_j)\setminus f_{j,i}}
\big[\bm C_{f_{j,k}\arrow j}^{(\ast)}\big]^{-1}\bm v^{(l-1)}_{f_{j,k}\arrow j}
 \bigg]\bigg\}.
\end{eqnarray}
Without loss of generality, define $\bm v^{(l)}$ as a vector containing all $\bm v^{(l)}_{f_{i,j}\arrow i}$
with ascending index\footnote{The order of $\bm v^{(l)}_{f_{i,j}\arrow i}$ arranged in $\bm v^{(l)}$
          in fact can be arbitrary as long as it does not change after the order is fixed.}
first on $i$ and then on $j$.
We can write (\ref{f2f2}) as
\begin{equation}\label{f2f23}
\bm v^{(l)}_{f_{i,j}\arrow i}
=\bm M_{i,j}\bm a_{i,j}
-\bm M_{i,j}\bm F_{i,j}\bm \Gamma_{i,j}\bm v^{(l-1)},
\end{equation}
where $\bm \Gamma_{i,j}$ is a block matrix containing $\big[\bm C_{f_{j,k}\arrow j}^{(\ast)}\big]^{-1}$ as component blocks such that (\ref{f2f23}) is satisfied.
Stacking (\ref{f2f23}) for all $i$ and $j$, and writing $\bm v^{(l)}\triangleq\big[(\bm v_x^{(l)})^T, (\bm v_y^{(l)})^T\big]^T$, where
$\bm v^{(l)}_x$ containing $\bm v^{(l)}_{f_{i,j}\arrow i}$ with $j=1$, while $\bm v^{(l)}_y$ containing the remaining part of $\bm v^{(l)}$, we obtain
\begin{equation} \label{vectoriter-x2}
\underbrace{\left[
\begin{array}{c }
\bm v_{x}^{(l)}
\\
{\bm v}_{y}^{(l)}
\end{array}
\right]}_{\triangleq\bm v^{(l)}}=
\underbrace{\left[
\begin{array}{cc}
\bm X & \bm Y
\\
\bm Q_1 & \bm Q_2
\end{array}
\right]}_{\triangleq\bm Q}
\underbrace{\left[
\begin{array}{c }
\bm v_{x}^{(l-1)}
\\
{\bm v}_{y}^{(l-1)}
\end{array}
\right]}_{\triangleq\bm v^{(l-1)}}
 +
\underbrace{\left[
\begin{array}{c }
\bm \xi_{x}
\\
{\bm \xi}_{y}
\end{array}
\right]}_{\triangleq\bm \xi}.
\end{equation}

On the other hand, putting $j=1$ into (\ref{f2f2}) and
notice that $\bm C_{f_1\arrow 1} =\bm 0$ and $\bm C^{(\ast)}_{1\arrow f_{i,1}} =\bm 0$ if $j=1$, we have
\begin{equation} \label{refm}
\bm v^{(l)}_{f_{i,1}\arrow i}
=
\big[\bm A_{i,1}^T \bm R_{i,1}^{-1}\bm A_{i,1}\big]^{-1}\bm A_{i,1}^T
\bm R_{i,1}^{-1}
(\bm r_{i,1}-\bm A_{1,i}\bm \omega_1).
\end{equation}
which shows that $\bm v^{(l)}_{f_{i,1}\arrow i}$  never changes with iteration number $l$.
Since $\bm v_x^{(l)}$ containing $\bm v^{(l)}_{f_{i,1}\arrow i}$ as components, $\bm v_x^{(l)}$ is fixed and independent of $l$.
Hence, we can write (\ref{vectoriter-x2}) equivalently as
\begin{equation} \label{vectoriter-2}
\underbrace{\left[
\begin{array}{c }
\bm v_{\textrm{fix}}
\\
{\bm v}_{y}^{(l)}
\end{array}
\right]}_{\bm v^{(l)}}=
\underbrace{\left[
\begin{array}{cc}
\bm I & \bm 0
\\
\bm Q_1 & \bm Q_2
\end{array}
\right]}_{\bm Q}
\underbrace{\left[
\begin{array}{c }
\bm v_{\textrm{fix}}
\\
{\bm v}_{y}^{(l-1)}
\end{array}
\right]}_{\bm v^{(l-1)}}
 +
\underbrace{\left[
\begin{array}{c }
\bm 0
\\
{\bm \xi_y}
\end{array}
\right]}_{\bm \xi}
\end{equation}
where $\bm v_{\textrm{fix}}=\bm v_{x}^{(l)}$ represents the stacked messages for $j=1$.
Notice that $[\bm Q_1 \bm Q_2]$ depends on $-\bm M_{i,j}\bm F_{i,j}\bm \Gamma_{i,j}$,
while $\bm \xi_y$ depends on $\bm M_{i,j}\bm a_{i,j}$.
It is obvious that $\bm Q_1$, $\bm Q_2$ and $\bm \xi_y$ are independent of iteration number $l$.
Next, we will show a property of $\bm Q_2$.

Since $\mathbb{E}_{\bm \omega_i,\bm{\omega_j},\bm n_{i,j}}\{\bm r_{i,j}\}=\mathbb{E}\{\bm A_{i,j}\bm \omega_i + \bm A_{j,i}\bm \omega_j + \bm n_{i,j}\}=\bm 0$,
and $[\bm C_{f_j\arrow j}]^{-1}=\bm 0$ with $j\neq 1$ for non-informative prior,
taking expectation on both sides of (\ref{vectoriter-2}), we have
\begin{equation} \label{vectoriter-3}
\underbrace{\left[
\begin{array}{c }
\bar{\bm v}_{\textrm{fix}}
\\
\bar{{\bm v}}_{y}^{(l)}
\end{array}
\right]}_{\triangleq\bar{\bm v}^{(l)}}=
\underbrace{\left[
\begin{array}{cc}
\bm I & \bm 0
\\
\bm Q_1 & \bm Q_2
\end{array}
\right]}_{\bm Q}
\underbrace{\left[
\begin{array}{c }
\bar{\bm v}_{\textrm{fix}}
\\
\bar{{\bm v}}_{y}^{(l-1)}
\end{array}
\right]}_{\triangleq\bar{\bm v}^{(l-1)}}
\end{equation}
or equivalently
\begin{equation}\label{transmean}
\bar{\bm v}^{(l)}=\bm Q^{l}\bar{\bm v}^{(0)},
\end{equation}
where $\bar{x}$ denotes the expectation of $x$.
Since there is always a positive value $\eta$, satisfying $\eta>\sum_{i\neq j}|[\bm Q]_{i,j}|$ for all $i$,
we have $\eta\bm I+\bm Q$ is strictly diagonally dominant and then $\eta\bm I+\bm Q$ is nonsingular \cite{MatrixHorn}.
Hence, arbitrary initial value $\bar{\bm v}^{(0)}$ can be expressed in terms of the eigenvectors of $\eta\bm I+ \bm Q$
as $\bar{\bm v}^{(0)} = \sum_{d=1}^{D}c_{d}\bm q_{d}$,
where $\bm q_1$, $\bm q_2$,$\cdots$, $\bm q_D$ are the eigenvectors of $\eta\bm I+\bm Q$.
Since the eigenvectors of $\eta\bm I+\bm Q$ is the same as that of $\bm Q$, and the eigenvalues of $\eta\bm I+\bm Q$ are $\eta+\lambda_d$ ($1\leqslant d \leqslant D$), where $\lambda_d$ is the eigenvalue of $\bm Q$, we have
\begin{eqnarray} \label{eigen}
\bar{\bm v}^{(l)}=
\bm Q^{l}\bar{\bm v}^{(0)}
= \sum_{{d}=1}^{D} {c_{d}}{\lambda_{d}}^{l}\bm q_d.
\end{eqnarray}
Without loss of generality, suppose ${\lambda}_{d}$ are arranged in  descending order as below
\begin{equation}
|{\lambda}_1|\geq|{\lambda}_2|\geq  \cdots \geq|{\lambda}_{D}|.
\end{equation}
Let the eigenvalue with the largest magnitude has a multiplicity of  $d_0$.
Then ${{\lambda}_{{d}}}/{{\lambda}_{1}}<1$ for ${d}>{d}_0$
and $({{\lambda}_{{d}}}/{{\lambda}_{1}})^l=0$ if $l$ is large enough.
We then obtain
\begin{eqnarray} \label{eigen2}
\bar{\bm v}^{(l)}
= {\lambda}_1^{l} \sum_{{d}=1}^{{d}_0}c_{d}\bm q_{d},
\end{eqnarray}
for large $l$.
Taking expectation on  (\ref{refm}), we have $\bar{\bm v}^{(l)}_{f_{i,1}\arrow i} = -\big[\bm A_{i,1}^T \bm R_{i,1}^{-1}\bm A_{i,1}\big]^{-1}\bm A_{i,1}^T\bm R_{i,1}^{-1}\bm A_{1,i}\bm \omega_1$.
It is obvious that $\bar{\bm v}^{(l)}_{f_{i,1}\arrow i}$ never change with the iteration number $l$.
Hence,
the first element of $\bar{{\bm v}}^{(l)}$ is $\bar{{\bm v}}^{(l)}(1)\triangleq \gamma_c$, which is a constant,
and according to (\ref{eigen2}), we have
${\lambda}_1^{l} = \frac{\gamma_c}{\sum_{{d}=1}^{{d}_0} c_{d}\bm q_{d}(1) }$
for $l$ large enough.
Substituting it back to (\ref{eigen2}) yields
\begin{equation}
\bar{\bm v}^{(l)} = \frac{\gamma_c\sum_{{d}=1}^{{d}_0}c_{d}\bm q_{d}}{\sum_{{d}=1}^{{d}_0} c_{d}\bm q_{d}(1) }.
\end{equation}
It is obvious that $\bar{\bm v}^{(l)} $ does not change when $l$ is large enough, and
therefore, $\bar{\bm v}^{(l)}$ in (\ref{transmean}) converges.
Since (\ref{vectoriter-3}) and (\ref{transmean}) are equivalent, $\bar{{\bm v}}_{y}^{(l)}$ in  (\ref{vectoriter-3}) also converges.
With iteration equation in (\ref{vectoriter-3}) rewritten as
\begin{equation} \label{vectoriter2-2}
\bar{\bm v}_{y}^{(l)}
=\bm Q_2\bar{\bm v}_{y}^{(l-1)}
+\bm Q_1
\bar{\bm v}_{\textrm{fix}},
\end{equation}
and since (\ref{vectoriter2-2}) converges, the spectrum radius $\rho({\bm Q}_2)<1$ \cite{Demmel}.

Now rewriting (\ref{vectoriter-2}) as
\begin{equation} \label{vectoriter1-2}
{\bm v}_{y}^{(l)}
=\bm Q_2{\bm v}_{y}^{(l-1)}
+\bm Q_1\bm v_{\textrm{fix}}
 +
{\bm \xi}_y.
\end{equation}
With $\bm Q_1\bm v_{\textrm{fix}}$ being a constant vector, and $\rho(\bm Q_2)<1$, we also have (\ref{vectoriter1-2}) converges.
Thus, the sequence $\bm v^{(l)}_{f_{i,j}\arrow i}$ in (\ref{f2f2})
is convergent for any initial vectors $\bm v^{(0)}_{f_{i,j}\arrow i}$ \cite{Demmel}.
Finally, with $\bm \mu_i^{(l)}$ defined in (\ref{beliefu}), since
$ \bm P_{i}^{(l)}$, $\bm C_{f_{i,j}\arrow i}^{(l)}$ and $\bm v^{(l)}_{f_{i,j}\arrow i}$
converge, we can draw the conclusion that the
vector sequence $\{\bm\mu_i^{(1)}, \bm\mu_i^{(2)}, \ldots\}$
converges.
\QEDA

Although Theorem 2 states that the proposed iterative estimation for $ \{{\bm \omega}_i\}_{i\in \mathcal{V}}$ converges to a fixed point $\{\bm \mu_i^{\ast}\}_{i\in \mathcal{V}}$,
We still need to answer the important question that
how accurate the converged $\{\bm \mu_i^{\ast}\}_{i\in \mathcal{V}}$ is?

{\textbf{Theorem 3.}}
The BP message mean vector $\bm \mu_i^{\ast}$ converges to the optimal estimates
$ \hat{\bm \omega}_i^{\text{MMSE}}$.
Furthermore, for the non-informative prior of $\bm \omega=[{\bm \omega}^T_2,\ldots, \bm \omega^T_K]^T$,
the estimation MSE of $\bm \mu^{\ast}=[(\bm \mu_2^{\ast})^T,\ldots, (\bm \mu_K^{\ast})^T]^T$ asymptotically
(in high SNR or large training length $N$ or both) equals the centralized CRB of $\bm \omega$:
\begin{equation}\label{cfoCRB}
\textrm{CRB}(\bm \omega)=\big(\bm A^T\bm R^{-1}\bm A\big)^{-1},
\end{equation}
where $\bm A$ is obtained from stacking (\ref{linearCFO}) into the form of $\bm r=\bm A\bm \omega+ \bm n$, with $\bm r$ being a vector containing $\bm r_{i,j}$ with ascending index first on $i$ and then on $j$;
$\bm R$ is a block diagonal matrix with $\bm R_{i,j}$ as block diagonal and with the same order as $\bm r_{i,j}$ in $\bm r$.

\noindent  \textit{Proof}:
Since the joint posterior distribution in (\ref{joint}) is multivariate Gaussian and it is known that if Gaussian BP converges, the mean of the beliefs computed by BP equals the mean of the marginal posterior distribution in (\ref{MMSE}), i.e., $ \bm \mu_i^{\ast}=\hat{\bm \omega}^{\text{MMSE}}_i $ \cite{DiagnalDominant}.

Notice that  stacking (\ref{MMSE}) into $\hat{\bm \omega}^{\text{MMSE}}=[(\hat{\bm \omega}^{\text{MMSE}}_2)^T,\ldots, (\hat{\bm \omega}^{\text{MMSE}}_K)^T]^T$ gives
\begin{equation} \label{centralizedMMSE}
\hat{\bm \omega}^{\text{MMSE}}
=   \int...\int
{\bm \omega}
p\big(\bm \omega_1,\bm \omega_2,\ldots,\bm \omega_K |{\{{\bm r}_{i, j}\}}_{\{i,j\}\in \mathcal{E}}\big)   d\bm \omega_2\ldots d\bm \omega_{K}.
\end{equation}
It is obvious that $\bm \mu^{\ast}=\big[(\bm\mu_2^*)^T, \ldots , (\bm\mu_K^*)^T\big]^T$ equals the centralized joint MMSE estimator $\hat{\bm \omega}^{\text{MMSE}}$.
Putting (\ref{joint}) into (\ref{centralizedMMSE}) and in case of non-informative prior, $ \hat{\bm\omega}^{\textrm{MMSE}}$ is the mean of the joint likelihood function.  Since the mean and maximum of a Gaussian distribution are the same,
therefore, $\bm \mu^{\ast}$ equals the centralized joint maximum likelihood estimator and it asymptotically approaches the centralized CRB \cite{Kay}.
Finally, stacking (\ref{linearCFO}) into the form
\begin{equation}\label{blocklinear}
\bm r= \bm A\bm \omega + \bm n,
\end{equation}
where $\bm r$ is a vector containing $\bm r_{i,j}$ with ascending indexes first on $i$ and then on $j$;
and $\bm n$ containing $\bm n_{i,j}$ with the indexes $i$, $j$ ordered in the same way as in $\bm r$.
Since $\bm n \sim \N(\bm n;\bm0,\bm R)$,
where $\bm R$ is a block diagonal matrix with $\bm R_{i,j}$ as block diagonal and with the same order as $\bm r_{i,j}$ in $\bm r$,  and (\ref{blocklinear}) is a standard linear model, the CRB for $\bm \omega$ is given by $\textrm{CRB}(\bm \omega)=\big(\bm A^T\bm R^{-1}\bm A\big)^{-1}$ \cite{Kay}.
\QEDA
\section{Simulation Results}\label{Section 5}
This section presents numerical results to assess the performance of the proposed algorithm.
Estimation MSE are presented for
CFO estimation over the whole random network, which consists of
$14$ nodes randomly located in a $[0, 100]\times[0, 100]$ area.
The communication range for each node is $38$.
In each trial, CFO of each antenna on each node (except node $1$ where CFO is zero)
is generated independently and is
uniformly distributed in the range $2\pi[-0.05,0.05]$.
It is assumed that we do not have prior information on the distribution of CFOs, so we set
$p(\bm\omega_i)= \N(\bm\omega_i;\bm 0, +\infty\bm I)$, $i=2,\ldots,14$.
Besides, the channel between each pair of nodes is Rayleigh flat-fading.
The relative CFOs and channels are first estimated based on the algorithm in\cite{Pham}, with training length $N$.  Then the BP algorithm is executed for network-wide CFOs estimation and compensation.
$5000$ simulation runs were performed to obtain the average performance for each point in the figures.

First, consider the network shown in Fig. \ref{topology} and
each node equipped with two antennas.
We employ training with length $N=16$ for relative CFOs estimation and the SNR during training stage in each node is $30\text{dB}$.
Fig. \ref{converge} shows the sum MSE\footnote{Sum MSE over the two antennas.}
of $\bm\omega_i$ for $i=\{3, 6, 2\} $ as a function of BP iteration number $l$.
It can be seen that the MSEs decrease quickly and touch the corresponding CRBs in only a few iterations.

Fig. \ref{average} shows the average sum MSE of $ \{{\bm \omega}_i\}_{i\in \mathcal{V}}$ versus SNRs for different training length $N$.
The network is randomly generated within the $[0, 100]\times[0, 100]$ area in each trial, and each node is equipped with $2$ antennas.
As shown in the figure, the MSEs of proposed distributed algorithm achieve the best performance as the MSEs touch the corresponding CRBs.
This verifies Theorem 3.
Furthermore, with increasing $N$,  $\bm R_{i,j}$ touches CRB of the relative CFO  $\bm B^{\{i,j\}}_{\bm{\epsilon}}(\{{{\bm{\epsilon}}}_{k}\}_{k=1}^{N_j},\{{{\bm{h}}}_{k}\}_{k=1}^{N_j})$ at lower SNR, and
thus the estimation MSEs of $\bm \omega_i$ achieves the corresponding CRBs earlier.

Finally, we compare the performance of the proposed algorithm with that of the D-FLL approach in \cite{Spagnolini}, where CFOs are adjusted to an average common value in each iteration based on consensus principle.
The estimation error of consensus method at the $l^{th}$ iteration is measured by
the total mean-square deviation of the individual variables from their average, which is
\begin{equation}
\text{MSE}^{(l)}_{\text{consensus}}=\frac{1}{K}\sum_{i=1}^{K}\mathbb{E}\bigg\{\big\|{\bm\mu}^{(l)}_i- \frac{1}{K}\sum_{i=1}^{K}{\bm\mu}^{(l)}_i \big\|^2\bigg\}.
\end{equation}
On the other hand, the proposed algorithm estimates the absolute CFO values, therefore, the network estimation MSE at the $l^{th}$ iteration is
\begin{equation}
\text{MSE}^{(l)}_{\text{BP}}=\frac{1}{K-1}\sum_{i=2}^{K}\mathbb{E}\bigg\{\big\|{\bm\mu}^{(l)}_i- \bm \omega_i\big\|^2\bigg\}.
\end{equation}

We consider the  $14$ nodes randomly located within the $[0, 100]\times[0, 100]$ area and
for fair comparison with D-FLL, each node is equipped with a single antenna.
For consensus method,  pilots of length $16$ are transmitted by each node in each iteration.
However, for the proposed method, $16$ pilots are used only in the relative CFOs estimation at the initial phase.
The convergence performance of the two algorithms at different SNRs are shown in Fig. \ref{consensus}.
It is apparent that convergence speed of the consensus algorithm decreases with SNRs, and in general takes several {hundreds of} iterations to converge.
For example, at $\textrm{SNR}=5$dB, around $800$ iterations is required.
Together with the fact that  a training of length $16$ is being transmitted by each node in each iteration, this is a huge burden to the network, and also causing long delay in the synchronization process.
On the other hand,
the proposed method requires only a few iterations of message exchange of two real numbers (mean and variance) to
approach the corresponding CRBs.

\section{Conclusions} \label{Section 6}
In this paper, a fully distributed CFOs  estimation
algorithm for  cooperative and distributed networks was proposed.
The algorithm is based on BP and is easy to be implemented by exchanging limited
amount of information between neighboring nodes.
Therefore, it has low overhead and is scalable with network size.
Furthermore, it was shown analytically that the proposed distributed algorithm converges
to the optimal solution with estimation MSE coincides with the centralized CRB
asymptotically regardless of the network topology.
Simulation results showed that the MSE of the proposed method touches the CRB within only a few
iterations.

\newpage
\begin{figure}[t]
  \centering
\mbox{\subfigure[Distributed beamforming networks]{\epsfig{figure=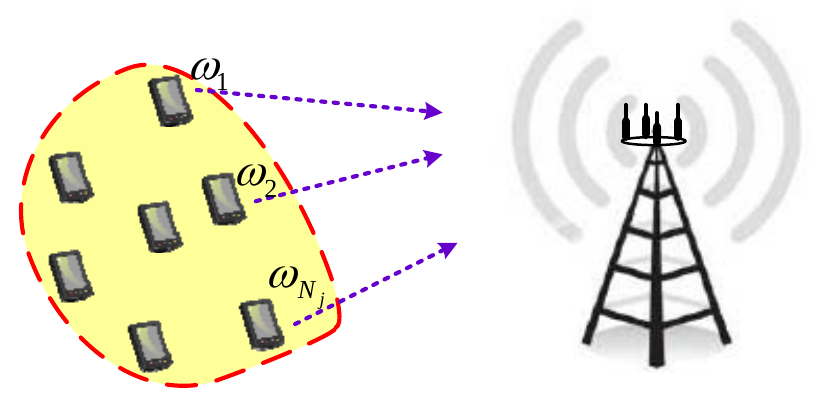,width=3in}}\label{DisBeam}
\qquad  \qquad
      \subfigure[Multi-cell cooperative networks]{\epsfig{figure=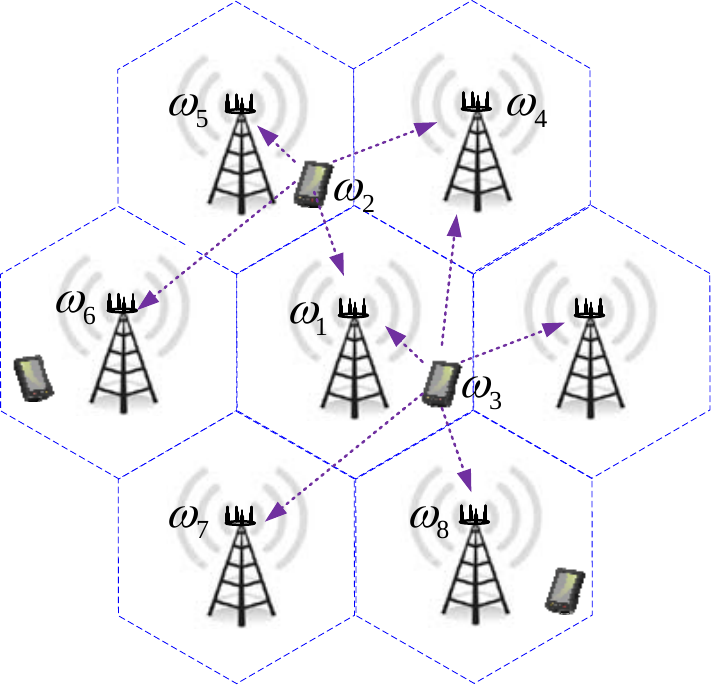,width=2in}}\label{MultiCell} }
\mbox{\subfigure[A 3-tier HetNet with macro, pico and femto cells.]{\epsfig{figure=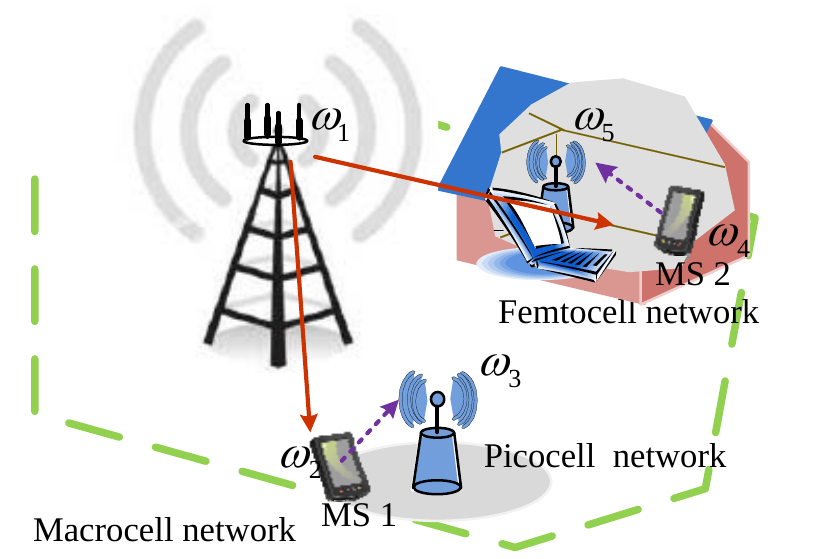,width=2.6in}}\label{HetNets}       }
\caption{Scenarios that need network-wide frequency synchronization}
  \label{fig:SubF}
\end{figure}


\begin{figure}[t]
\centering
\epsfig{file=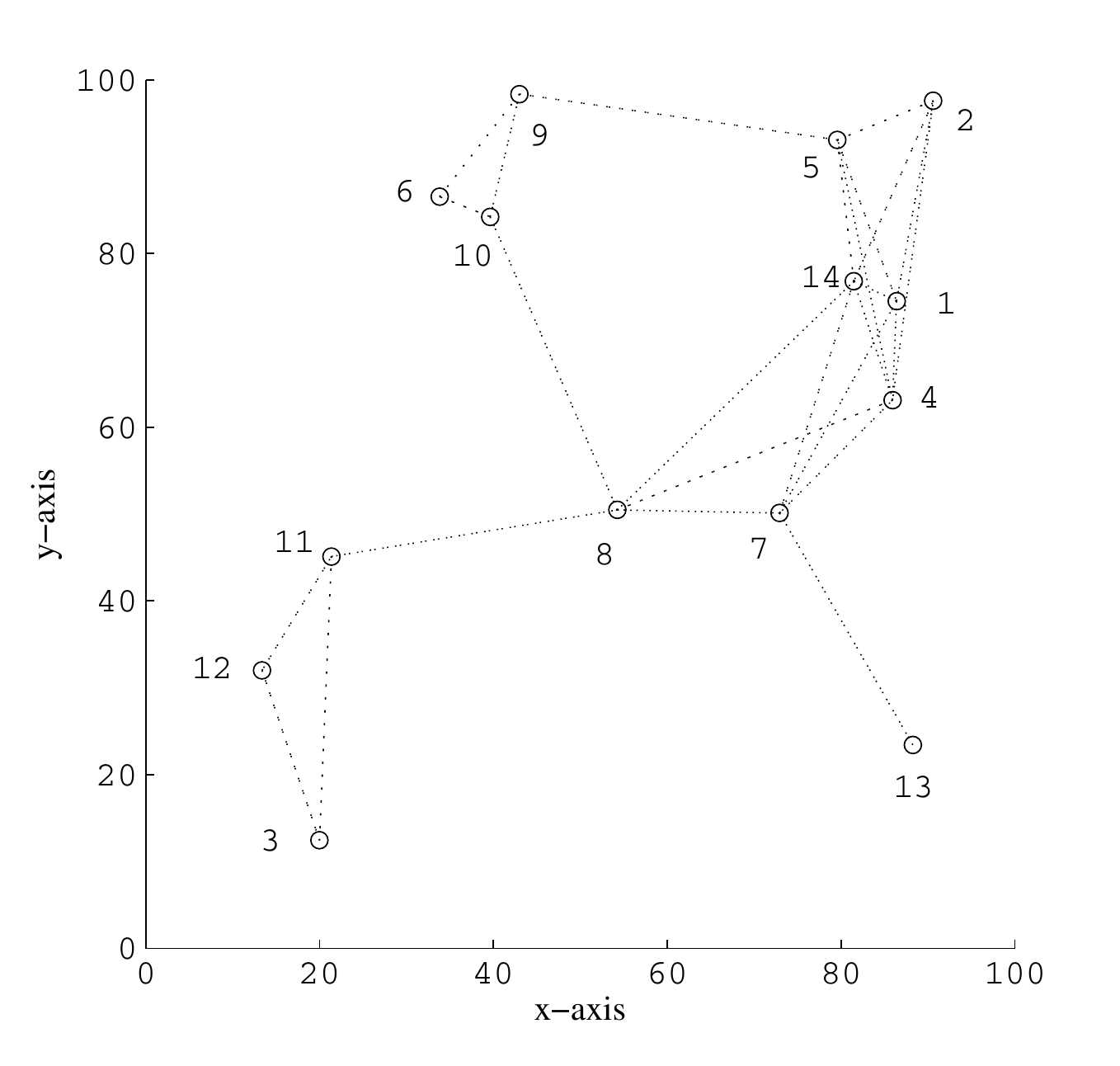, width=3in}
\caption{An example of arbitrary network topology with 14 nodes.}
\label{topology}
\end{figure}

\begin{figure}[t]
\centering
\epsfig{file=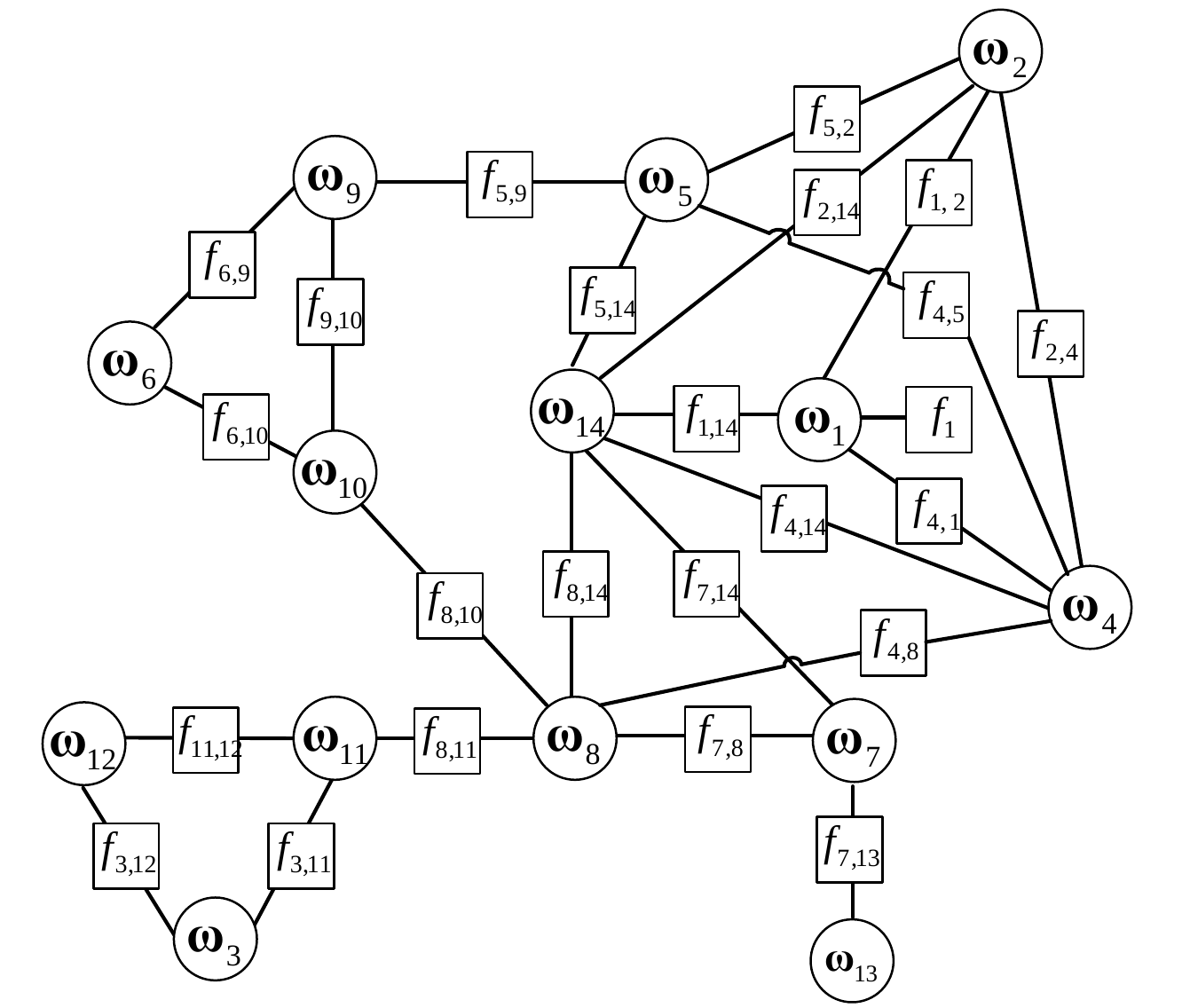, width=3in}
\caption{The factor graph representation of the network in Fig. 2.}
\label{FG}
\end{figure}

\begin{figure}[t]
\centering
\epsfig{file=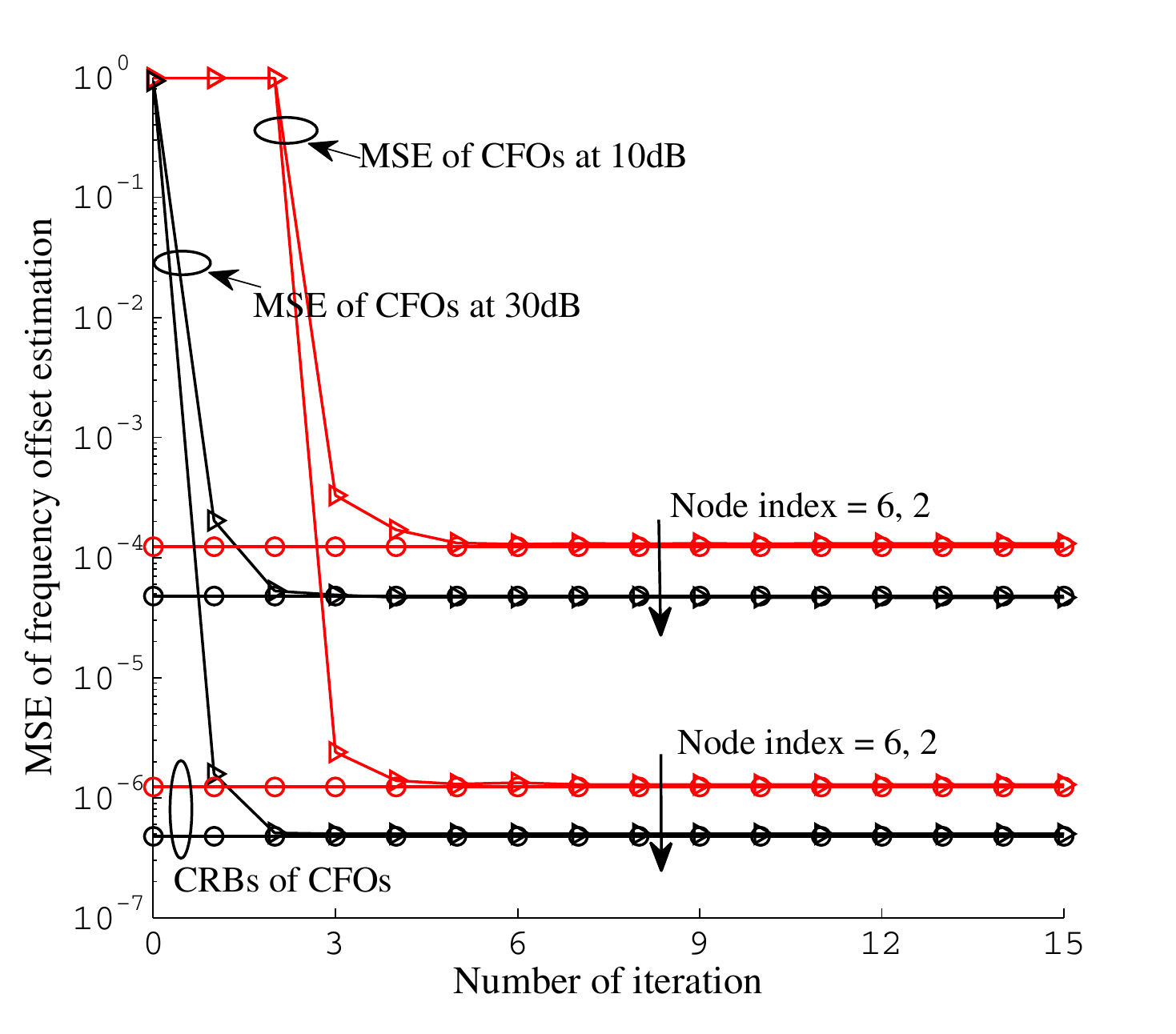, width=3.2in}
\caption{Convergence performance of the proposed algorithm at different nodes.}
\label{converge}
\end{figure}

\begin{figure}[t]
\centering
\epsfig{file=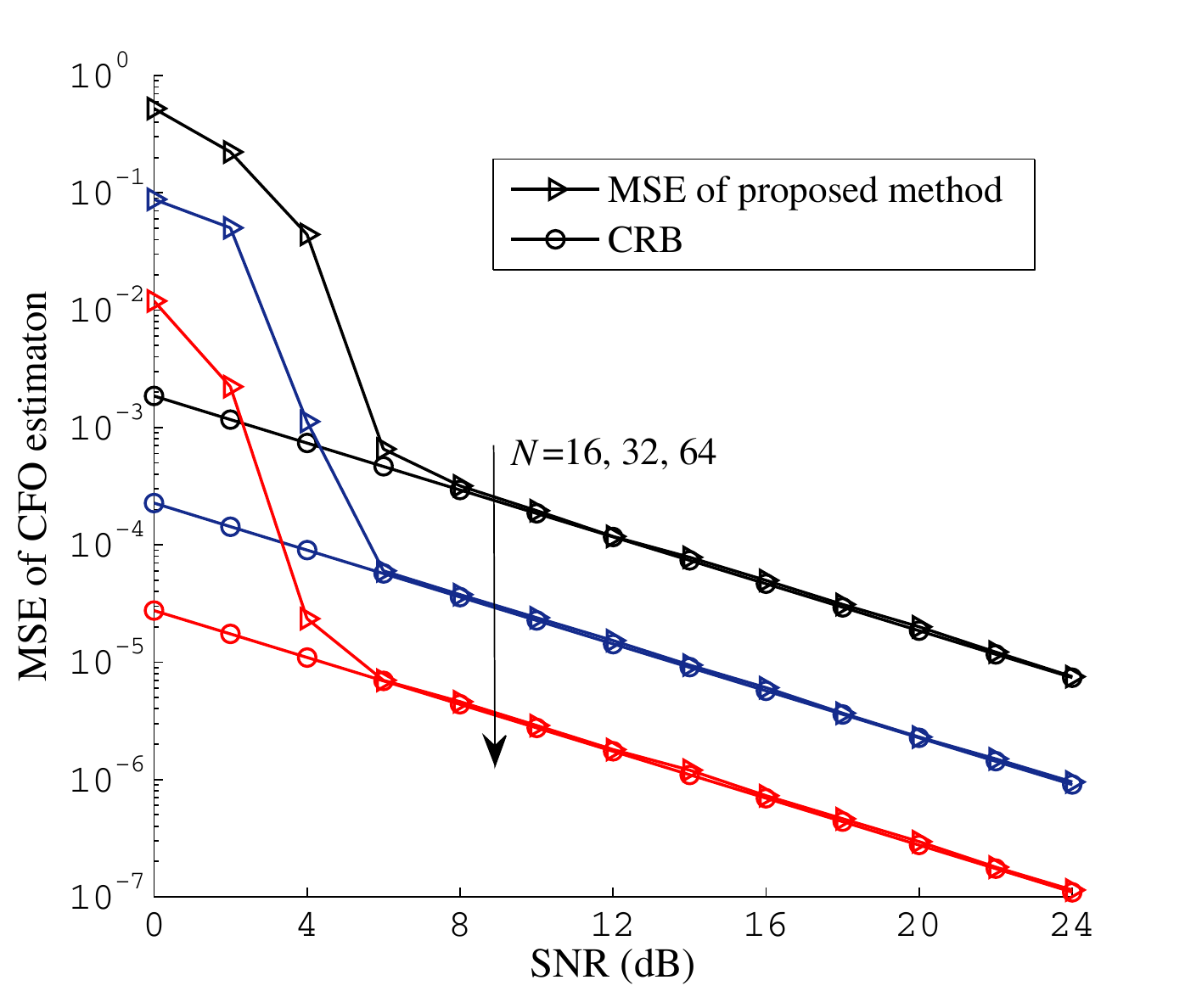, width=3.5in}
\caption{MSE of CFOs $ \{{\bm \omega}_i\}_{i\in \mathcal{V}}$ averaged over the whole network with respect to SNRs.}
\label{average}
\end{figure}

\begin{figure}[t]
\centering
\epsfig{file=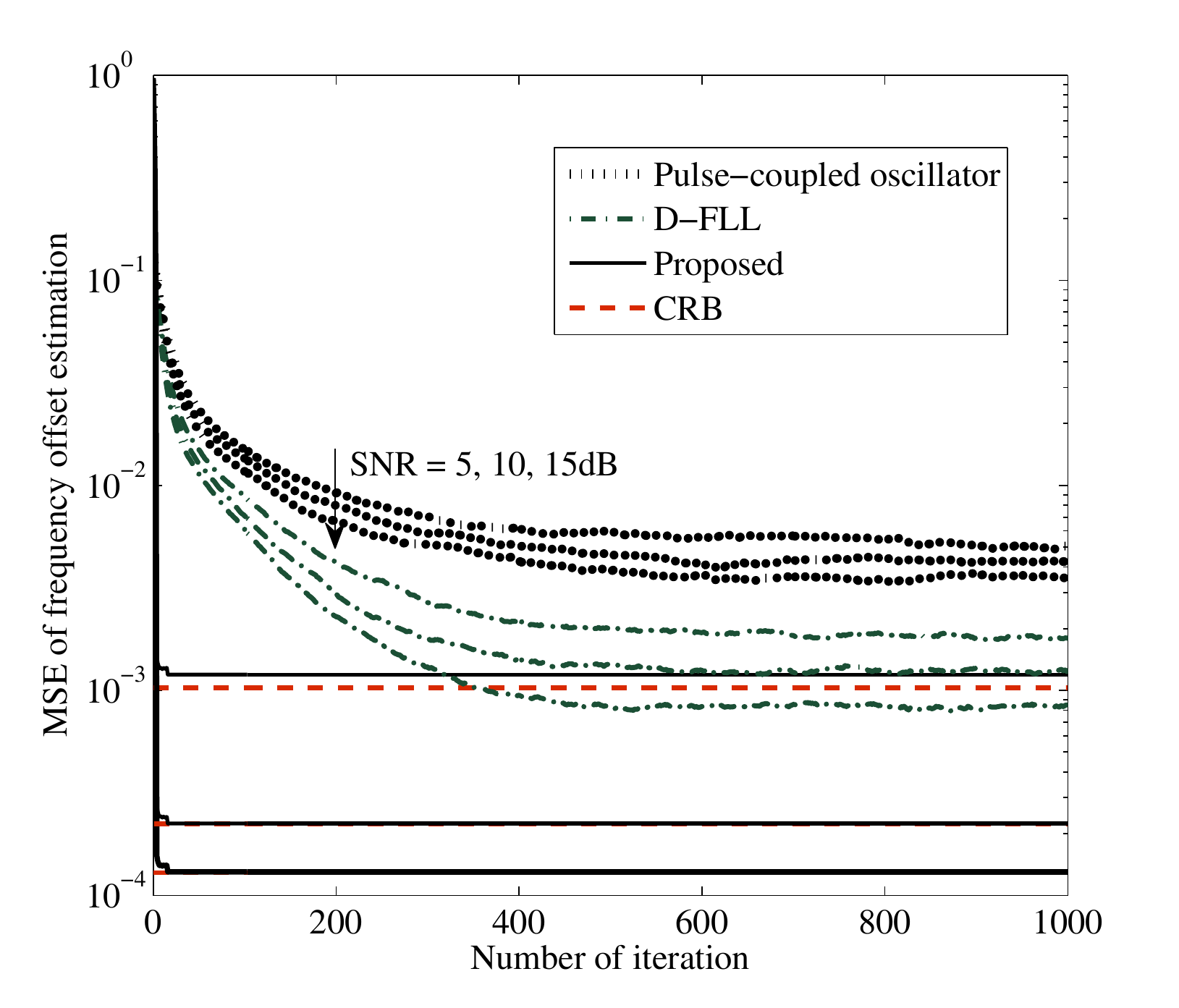, width=3.7in}
\caption{Convergence of the proposed method and consensus method in \cite{Spagnolini} in single antenna case.}
\label{consensus}
\end{figure}

\end{document}